**Evidence for multiple magma ocean outgassing and atmospheric loss episodes from mantle noble gases**

Jonathan M. Tucker* and Sujoy Mukhopadhyay

Department of Earth and Planetary Sciences, Harvard University, Cambridge, MA 02138, USA

*Corresponding author: jtucker@fas.harvard.edu, 617-496-4475

**Abstract**

The energy associated with giant impacts is large enough to generate global magma oceans during Earth's accretion. However, geochemical evidence requiring a terrestrial magma ocean is scarce. Here we present evidence for at least two separate magma ocean outgassing episodes on Earth based on the ratio of primordial $^3$He to $^{22}$Ne in the present-day mantle. We demonstrate that the depleted mantle $^3$He/$^{22}$Ne ratio is at least 10 while a more primitive mantle reservoir has a $^3$He/$^{22}$Ne ratio of 2.3 to 3. The $^3$He/$^{22}$Ne ratios of the mantle reservoirs are higher than possible sources of terrestrial volatiles, including the solar nebula ratio of 1.5. Therefore, a planetary process must have raised the mantle's $^3$He/$^{22}$Ne ratio.

We show that long-term plate tectonic cycling is incapable of raising the mantle $^3$He/$^{22}$Ne ratio and may even lower it. However, ingassing of a gravitationally accreted nebular atmosphere into a magma ocean on the proto-Earth explains the $^3$He/$^{22}$Ne and $^{20}$Ne/$^{22}$Ne ratios of the primitive mantle reservoir. Increasing the mantle $^3$He/$^{22}$Ne ratio to a value of 10 in the depleted mantle requires at least two episodes of atmospheric blow-off and magma ocean outgassing associated with giant impacts during subsequent terrestrial accretion. The preservation of a low $^3$He/$^{22}$Ne ratio in a primitive reservoir sampled by plumes suggests that the later giant impacts, including the Moon-forming giant impact, did not generate a whole mantle magma ocean.

Atmospheric loss episodes associated with giant impacts provide an explanation for Earth's subchondritic C/H, N/H, and Cl/F elemental ratios while preserving chondritic isotopic ratios. If so, a significant proportion of terrestrial water and potentially other major volatiles were accreted prior to the last giant impact, otherwise the fractionated elemental ratios would have been overprinted by the late veneer.

# 1. Introduction

Global magma oceans are thought to be a natural result of the highly energetic giant impact phase of planetary accretion (Abe and Matsui, 1985; Benz and Cameron, 1990; Canup, 2008; Elkins-Tanton, 2012; Sasaki and Nakazawa, 1986; Stevenson, 1987; Tonks and Melosh, 1993). The best evidence for a terrestrial magma ocean is the relative concentrations of siderophile elements in the mantle, which suggest core formation occurred when at least part of the Earth's mantle was molten (Li and Agee, 1996; Righter et al., 1997; Rubie et al., 2007). However, elements that trace silicate differentiation do not show strong signatures for a magma ocean. For example, Kato et al. (1988) and Ringwood (1990) argued that refractory lithophile trace elements should have been highly fractionated by magma ocean crystallization and that fractionation is not observed in Hadean zircons, which preserve chondritic ratios. On the other hand, magma ocean crystallization models have been used to explain incongruent Sm-Nd and Lu-Hf isotope systematics in Archean rocks (Caro et al., 2005; Rizo et al., 2011) as well a slightly superchondritic mantle Ca/Al ratio (Walter et al., 2004). Geochemical evidence for a magma ocean may be difficult to find in the present-day mantle, as crustal recycling and mantle mixing throughout Earth's history may have erased the chemical fractionations of the lithophile elements produced by magma ocean crystallization.

The noble gases may provide unique information on the occurrence of magma oceans, as magma ocean ingassing or outgassing will create large fractionations in noble gas elemental ratios due to their different solubilities in magma. The present-day budgets of nonradiogenic Ar, Kr, and Xe in the mantle are dominated by recycled air (Holland and Ballentine, 2006; Kendrick et al., 2011; Mukhopadhyay, 2012; Pető et al., 2013; Sumino et al., 2010; Tucker et al., 2012), so possible magma ocean degassing signatures involving Ar, Kr, and Xe would have been overprinted by subduction. On the other hand, He and Ne are not recycled back into the mantle in significant quantities, and may preserve ancient magma ocean signatures (Harper and Jacobsen, 1996; Honda and McDougall, 1998; Porcelli et al., 2001; Shaw et al., 2001; Yokochi and Marty, 2004).

Honda and McDougall (1998) observed that the mantle $^3$He/$^{22}$Ne ratio, as determined from measurements of mid-ocean ridge basalts (MORBs) and ocean island basalts (OIBs), was twice the solar value, and suggested that this fractionation resulted from the higher solubility of He compared to Ne during magma ocean outgassing. Ingassing, i.e. dissolution of volatiles from an accreted nebular atmosphere into a magma ocean (Harper and Jacobsen, 1996; Mizuno et al., 1980), could also have the same effect on the mantle $^3$He/$^{22}$Ne ratio. However, magma ocean equilibration (ingassing or outgassing) has been regarded as a speculative explanation of the high mantle $^3$He/$^{22}$Ne ratios (e.g., Graham, 2002). Yokochi and Marty (2004) advocated magma ocean ingassing based on observations of $^{20}$Ne/$^{22}$Ne ratios of >13.0 in the Kola plume, a Ne isotopic composition similar to the solar wind. The authors also noted that the high $^3$He/$^{22}$Ne ratio (~7.9) observed in the Kola plume and in present-day MORBs requires additional fractionation beyond that capable through magma ocean ingassing. They speculated that the fractionation mechanism was fluid-melt partitioning during high pressure melt segregation and fluid phase formation in the deep mantle. However, it is unclear whether fluid phases would nucleate at the P-T conditions of the deep mantle.

Here we present evidence for ancient magma ocean degassing based on high $^3$He/$^{22}$Ne ratios in the depleted mantle that are more extreme than the average MORB value. To establish the $^3$He/$^{22}$Ne ratio of the depleted mantle, we characterize how $^3$He/$^{22}$Ne variations are linked to lithophile (Pb, Sr, Nd) and noble gas isotope ratios in very depleted to enriched MORBs using published data from the equatorial Atlantic (Agranier et al., 2005; Schilling et al., 1994; Tucker et al., 2012). We then show that magma ocean degassing is the most likely explanation for the large $^3$He/$^{22}$Ne fractionation observed in the present-day mantle. We demonstrate that extraction of He and Ne from the mantle through partial melting and plate subduction associated with plate tectonics either has a negligible effect or decreases the mantle $^3$He/$^{22}$Ne ratio. We argue that the magnitude of the $^3$He/$^{22}$Ne fractionation requires atmospheric loss followed by solubility-controlled degassing of at least two separate magma oceans during the giant impact phase of Earth's accretion.

## 2. Determining MORB mantle source $^3$He/$^{22}$Ne ratios

Direct measurements of $^3$He/$^{22}$Ne ratios in mantle-derived basalts are not likely to represent the mantle value because the ratio can be changed by magmatic degassing during eruption and ubiquitous shallow-level air contamination. We therefore calculate the mantle source $^3$He/$^{22}$Ne ($^3$He/$^{22}$Ne$_m$) ratio from measured He and Ne isotope ratios in basalts, after Honda and McDougall (1998) and Porcelli and Ballentine (2002) (Method 1):

$$^3\text{He}/^{22}\text{Ne}_m = \frac{^{21}\text{Ne}/^{22}\text{Ne}_E - ^{21}\text{Ne}/^{22}\text{Ne}_i}{^4\text{He}/^3\text{He}_{meas} - ^4\text{He}/^3\text{He}_i} \times (^4\text{He}/^{21}\text{Ne})^*_{production} \qquad (1)$$

$^{21}$Ne/$^{22}$Ne$_E$ is the mantle source $^{21}$Ne/$^{22}$Ne ratio corrected for shallow-level air contamination by extrapolation of measured $^{21}$Ne/$^{22}$Ne ratios to the mantle $^{20}$Ne/$^{22}$Ne ratio of 12.5 (Tucker et al., 2012). $^{21}$Ne/$^{22}$Ne$_i$, the initial $^{21}$Ne/$^{22}$Ne ratio for the mantle, is 0.0313, corresponding to a $^{20}$Ne/$^{22}$Ne ratio of 12.5. The primordial He isotope ratio $^4$He/$^3$He$_i$ is 6024 (120 $R_A$), as observed in the Jovian atmosphere (Mahaffy et al. 1998), and ($^4$He/$^{21}$Ne)$^*_{production}$ is the production ratio of radiogenic $^4$He to nucleogenic $^{21}$Ne in the mantle.

We additionally calculate the $^3$He/$^{22}$Ne$_m$ ratio (Method 2) by first correcting the sample's measured $^3$He/$^{22}$Ne and $^4$He/$^{21}$Ne ratios for air contamination ($^3$He/$^{22}$Ne$_E$ and $^4$He/$^{21}$Ne$_E$; Fig. 1) and computing ($^4$He/$^{21}$Ne)* (where '*' refers to the mantle-derived radiogenic/nucleogenic species) from $^4$He/$^{21}$Ne$_E$. $^3$He/$^{22}$Ne$_E$ is corrected for magmatic degassing by the degree to which the sample's ($^4$He/$^{21}$Ne)* ratio is fractionated from the expected production ratio (e.g., Graham, 2002; see supplementary material). The two methods give comparable results for our MORB samples (Table 1).

In our calculations, we assume the MORB mantle $^{20}$Ne/$^{22}$Ne ratio is 12.5 (Ballentine et al., 2005; Holland and Ballentine, 2006; Raquin et al., 2008; Trieloff et al., 2000). If we instead assume a MORB mantle $^{20}$Ne/$^{22}$Ne ratio of 13.8, corresponding to the solar wind value (Grimberg et al., 2006), calculated $^3$He/$^{22}$Ne$_m$ ratios increase by ~50%, strengthening our conclusions. Additionally, we report $^3$He/$^{22}$Ne$_m$ ratios using two different determinations of ($^4$He/$^{21}$Ne)*$_{production}$: 2.2×10$^7$ (Yatsevich and Honda, 1997) and 2.8×10$^7$ (Ballentine and Burnard, 2002; Leya and Wieler, 1999). For consistency, we refer to the $^3$He/$^{22}$Ne$_m$ ratios calculated with the lower production ratio, noting this yields a lower bound on $^3$He/$^{22}$Ne$_m$ ratios. We further note that using $^4$He/$^3$He$_i$ ratios of ~2000–3000, corresponding to modern solar wind, or using a mantle $^{20}$Ne/$^{22}$Ne ratio of ~12.4, corresponding to highest measured values in our samples, has a negligible effect on the computed $^3$He/$^{22}$Ne$_m$ ratio.

## 3. The $^3$He/$^{22}$Ne ratio of the depleted mantle

The equatorial Atlantic MORBs are derived from a heterogeneous source with $^3$He/$^{22}$Ne ratios of 6.1 to 9.8 (Table 1). The most depleted MORBs are derived from a mantle with the highest $^3$He/$^{22}$Ne ratios and the more enriched MORBs (Schilling et al., 1994) are derived from a mantle with lower $^3$He/$^{22}$Ne ratios (Fig. 2). We note that an average MORB $^3$He/$^{22}$Ne$_m$ ratio of 10.2 ± 1.6 was calculated by Honda and McDougall (1998) and 8.8 ± 3.5 by Graham (2002). These values were calculated assuming a mantle $^{20}$Ne/$^{22}$Ne ratio of 13.8, and under this assumption, our samples range from 8.9–14.3. The average $^3$He/$^{22}$Ne$_m$ ratios of Honda and McDougall (1998) and Graham (2002) recalculated to $^{20}$Ne/$^{22}$Ne = 12.5 are 7.3 ± 1.2 and 6.1 ± 2.4, respectively, at the mid to low end of our range (Table 1; supplementary material).

Since $^3$He/$^{22}$Ne$_m$ ratios are strongly correlated with the lithophile isotopic composition of the MORBs (Fig. 2), the variations in $^3$He/$^{22}$Ne$_m$ ratios must have been produced by recent mixing between a depleted mantle with a high $^3$He/$^{22}$Ne ratio (≥~10) and an enriched mantle with a low $^3$He/$^{22}$Ne ratio. Specifically, intermediate values can be explained by mixing depleted mantle with primitive and recycled material (e.g., Schilling et al., 1994; Tucker et al., 2012; see supplementary material). Because the lithophile isotopic composition of the depleted samples defines the depleted end of the MORB array (Schilling et al., 1994), the depleted mantle $^3$He/$^{22}$Ne ratio must be at least 9.8 (Fig. 2). As the $^3$He/$^{22}$Ne$_m$ value of 9.8 is itself a lower limit for the most depleted sample (Section 2), in subsequent discussions we take the depleted mantle to have a $^3$He/$^{22}$Ne ratio of ≥~10.

Previous studies have attributed variations in $^3$He/$^{22}$Ne$_m$ ratios in mantle-derived rocks to either source heterogeneity (Coltice et al., 2011; Kurz et al., 2009; Moreira et al., 2001; Mukhopadhyay, 2012; Raquin and Moreira, 2009; Yokochi and Marty, 2004) or to fractionation resulting from partial melting and degassing of melts derived from a mantle with a uniform $^3$He/$^{22}$Ne ratio (Moreira and Allegre, 1998; Moreira et al., 2001; Sarda et al., 2000; Trieloff and Kunz, 2005; Hopp and Trieloff, 2008). However, the correlations between the $^3$He/$^{22}$Ne$_m$ and lithophile isotopic ratios in the MORB samples (Fig. 2) clearly demonstrate the existence of mantle domains with different $^3$He/$^{22}$Ne ratios. We further note that the correlations argue against source variations in $^3$He/$^{22}$Ne arising from fractionation due to fluid/melt partitioning of gases in the deep mantle (Yokochi and Marty, 2004).

## 4. $^3$He/$^{22}$Ne ratios in the present-day mantle and in planetary materials

Based on the mixing trends in Figure 2, the present-day depleted mantle has a $^3$He/$^{22}$Ne ratio of ≥10. In contrast, plumes tend to have lower $^3$He/$^{22}$Ne ratios (e.g. Füri et al., 2010; Graham, 2002; Honda and McDougall, 1998; Jackson et al., 2009; Kurz et al., 2009; Moreira et al., 2001; Mukhopadhyay, 2012; Raquin and Moreira, 2009; Yokochi and Marty, 2004). OIBs with the most primitive $^{21}$Ne/$^{22}$Ne ratios (e.g., Galapagos, Iceland), have $^3$He/$^{22}$Ne$_m$ ratios of 2.3 to 3 (Kurz et al., 2009; Mukhopadhyay, 2012; Raquin and Moreira, 2009; also see supplementary material).

The present-day mantle is also enriched in $^3$He relative to $^{22}$Ne compared to the sources that may have contributed primordial He and Ne to Earth (Fig. 3). These sources include the solar nebula ($^3$He/$^{22}$Ne = 1.46 ± 0.06), the implanted solar (Ne-B) component in gas-rich meteorites ($^3$He/$^{22}$Ne = 0.9 ± 0.1), the chondritic value ($^3$He/$^{22}$Ne ≤ 0.89 ± 0.11), as well as differentiated meteorites (Fig. 3; also see supplementary material).

The particular source(s) of He and Ne to the Earth are unclear, but must be consistent with Ne isotopes. For example, the observation of $^{20}$Ne/$^{22}$Ne ratios >12.9 in the Iceland and Kola plumes (Mukhopadhyay, 2012; Yokochi and Marty, 2004) implies a nebular origin for Ne, corresponding to a $^3$He/$^{22}$Ne of 1.46 ± 0.06. The $^3$He/$^{22}$Ne ratio of the primitive reservoir sampled by plumes is, therefore, fractionated from the nebular value by approximately a factor of 1.5 to 2. The MORB source $^{20}$Ne/$^{22}$Ne ratio of ~12.5 is similar to the Ne-B component observed in some primitive meteorites (Ballentine et al., 2005; Holland and Ballentine, 2006; Raquin et al., 2008), thought to represent implantation of solar wind into dust grains (Grimberg et al., 2006; Raquin and Moreira, 2009). Alternatively, the MORB source $^{20}Ne/^{22}Ne$ ratio of ~12.5 could have originated through a limited amount of recycling of atmospheric Ne ($^{20}Ne/^{22}Ne$ = 9.8) into a mantle with nebular Ne (Kendrick et al., 2011). In either case, the $^{3}He/^{22}Ne$ ratio of the material that contributed He and Ne to the MORB source must have been ≤1.46 ± 0.06 (Fig. 3). Therefore, the present-day depleted mantle $^{3}He/^{22}Ne$ ratio is fractionated by at least a factor of 6.5 from possible sources of terrestrial volatiles (Fig. 3). Additionally, because of mixing between the depleted mantle and the primitive reservoir over time, the depleted mantle $^{3}He/^{22}Ne$ ratio was likely higher than 10 in the past. Thus, a factor of 6.5 is a lower limit for the degree of $^{3}He/^{22}Ne$ fractionation of the depleted mantle.

## 5. Can plate tectonics generate reservoirs with high $^{3}He/^{22}Ne$ ratios?

We now address the puzzling observation: how has the Earth's interior acquired $^{3}He/^{22}Ne$ ratios higher than even the solar nebula? We investigate whether processes associated with the present style of plate tectonics can generate the high $^{3}He/^{22}Ne$ ratio in the depleted mantle and the difference between the depleted mantle and the primitive reservoir. Specifically, the mechanisms that could change the $^{3}He/^{22}Ne$ ratio are (i) partial melting that generates oceanic crust and a depleted mantle residue and (ii) recycling and mixing of tectonic plates.

### 5.1 Partial melting

Partial melting could raise or lower the He/Ne ratio of lavas if the two elements have different partition coefficients during mantle melting. Such fractionation has been advocated to explain the apparent difference in $^3$He/$^{22}$Ne$_m$ ratios between MORBs and OIBs under the assumption of a uniform mantle $^3$He/$^{22}$Ne ratio (Füri et al., 2010; Hopp and Trieloff, 2008; Moreira and Allegre, 1998; Sarda et al., 2000; Trieloff and Kunz, 2005). Hopp and Trieloff (2008) presented a hypothesis for generating a low $^3$He/$^{22}$Ne$_m$ ratio in OIBs compared to MORBs. They proposed that melting of the MORB source does not fractionate the $^3$He/$^{22}$Ne ratio, while low degree melting of mantle plumes produces melts with significantly lower $^3$He/$^{22}$Ne ratios than the source. Mixing of melts derived from the MORB source with the low degree melts from the plume source then generates the range of $^3$He/$^{22}$Ne$_m$ ratios measured in mantle-derived rocks.

Experimentally determined partition coefficients, however, rule out partial melting as the primary mechanism for producing the large range of observed $^3$He/$^{22}$Ne$_m$ ratios (Table 2; Brooker et al., 2003; Heber et al., 2007; Jackson et al., 2013). For example, generating the observed MORB/OIB fractionation of ~3–4× would require a degree of partial melting on the order of 0.01%, far too low for the generation of OIBs. Given a typical degree of melting of 1–10% for OIBs and MORBs, the $^3$He/$^{22}$Ne ratio of the melt would be different from the source by at most 0.3–3.6% compared to the large (>100%) variability observed (Figs. 2,3). Therefore, we conclude that variability in $^3$He/$^{22}$Ne$_m$ ratios reflects differences in the $^3$He/$^{22}$Ne ratio of the mantle and not fractionation produced through partial melting.

*5.2 Recycling and mixing of slabs*

The process of generating slabs and recycling them back to the mantle creates chemical heterogeneity in the mantle and has likely occurred since the Hadean (Coltice and Schmalzl, 2006; Harrison et al., 2005; Hopkins et al., 2008). Partial melting of the mantle generates the oceanic crust and depleted residue. The partial melt (oceanic crust) degasses, and the slab, comprised of the degassed crust and depleted residue, is recycled during subduction and mixed back into the mantle.

The partial melt will have the same $^3$He/$^{22}$Ne ratio as the mantle source (Section 5.1). Degassing of the partial melt prior to solidification into oceanic crust could then generate high $^3$He/$^{22}$Ne ratios in the curst as He is more soluble in the melt than Ne. However, recycling and mixing of oceanic crust back into the mantle could not have increased the mantle $^3$He/$^{22}$Ne ratio for a number of reasons. First, direct measurements reveal that old basaltic and gabbroic portions of the oceanic crust have $^3$He/$^{22}$Ne ratios of ~0 due to the ubiquitous introduction of atmospheric noble gases (Moreira et al., 2003; Staudacher and Allègre, 1988). Recycling and mixing of oceanic crust would, therefore, serve to *lower* the mantle $^3$He/$^{22}$Ne ratio over time. Second, if recycling of partially degassed oceanic crust introduced a high $^3$He/$^{22}$Ne ratio into the mantle, then HIMU (high-μ, where μ is the U/Pb ratio) OIBs, known to sample recycled oceanic crust (e.g., Cabral et al., 2013) and enriched MORBs from the equatorial Atlantic that likely sample recycled crust (Schilling et al., 1994) should have the highest $^3$He/$^{22}$Ne$_m$ ratios. However, the observations show the *opposite* to be true (Figs. 2,3). The enriched MORBs from the equatorial Atlantic have significantly lower $^3$He/$^{22}$Ne$_m$ ratios than the depleted MORBs, and HIMU OIBs from the Cook-Australs and Cameroon line have $^3$He/$^{22}$Ne$_m$ ratios of 4.6 ± 0.3 and 4.8 ± 1.1, respectively (Fig. 3; Barfod et al., 1999; Parai et al., 2009). Furthermore, trace element and isotopic compositions demonstrate that compared to enriched mantle domains, depleted mantle domains are less influenced by recycled oceanic crust. Therefore, partially degassed recycled oceanic crust cannot produce the high $^3$He/$^{22}$Ne$_m$ ratios observed in the depleted MORBs.

Instead, we explore whether mixing the residue of partial melting back into the mantle can generate the high $^3$He/$^{22}$Ne ratios. Because our goal is to understand how the mantle $^3$He/$^{22}$Ne ratio can increase to values ≥10, and because old oceanic crust has a $^3$He/$^{22}$Ne ratio of ~0, we will assume quantitative extraction of He and Ne from the oceanic crust during subduction in our simple model. For the simplest case where slabs (oceanic crust + residual mantle) are mixed instantaneously with the mantle, the concentration of a primordial isotope ($^3$He or $^{22}$Ne) in the mantle reservoir is:

$$C = C_0 e^{(f-1)N} \qquad (2)$$

where $C$ is the present-day concentration of the isotope in the mantle reservoir, $C_0$ the initial concentration, $f$ the fraction of gas retained in the mantle residue, and $N$ the number of reservoir masses processed through partial melting over 4.5 Ga (Gonnermann and Mukhopadhyay, 2009). The fraction of He and Ne in the residue depends on that element's bulk partition coefficient, $D$, and the extent of partial melting, $F$. Assuming batch melting, the concentration of the primordial isotope in the mantle as a function of $N$ is:

$$C = C_0 e^{\left(\frac{D}{F+D(1-F)}-1\right)N} \qquad (3)$$

To generate the largest $^3$He/$^{22}$Ne fractionation, we use the extremes in the ranges of He and Ne partition coefficients (Table 2) during partial melting of a mantle with 60% olivine, 20% clinopyroxene, 20% orthopyroxene and a global average $F$ of 0.05. We observe that the $^3$He/$^{22}$Ne ratio of the mantle changes by at most 13% even after 20 reservoir masses have been processed (Fig. 4). Hence, starting with a solar nebular $^3$He/$^{22}$Ne ratio of 1.46 ± 0.06 does not produce the present-day OIB source of ~2.3 to 3 and starting with the OIB value does not produce the present-day depleted mantle values of ≥10 (Fig. 4). Furthermore, compared to fractional or dynamic melting, batch melting produces the largest effect on the mantle $^3$He/$^{22}$Ne ratio. For example, using the highest partition coefficients within the 1σ ranges (Table 2) and for a 1% partial melt, batch melting produces residues depleted in He and Ne by 2 orders of magnitude, while fractional melting produces residues depleted by 12–13 orders of magnitude. As a result, residues produced by fractional melting cannot affect the mantle's $^3$He/$^{22}$Ne ratio. Overall, our simple model demonstrates that recycling of slabs and extracting He and Ne from the mantle through partial melting cannot significantly increase the mantle $^3$He/$^{22}$Ne ratio.

## 6. Generating high $^3$He/$^{22}$Ne ratios via magma ocean ingassing and outgassing

While extraction of He and Ne from the mantle through partial melting and subsequent recycling of plates cannot significantly increase the mantle $^3$He/$^{22}$Ne ratio, ingassing or outgassing of a magma ocean would produce a mantle with a higher $^3$He/$^{22}$Ne ratio. The higher magmatic solubility of He compared to Ne results in preferential partitioning of Ne into the atmosphere and an increase in the $^3$He/$^{22}$Ne ratio of the mantle. In an equilibrium between a magma ocean and an atmosphere, the magma $^3$He/$^{22}$Ne ratio can be increased by up to a factor of the He/Ne solubility ratio ($S_{He}/S_{Ne}$) relative to the atmosphere. Previous suggestions of a magma ocean based on He/Ne studies (Honda and McDougall, 1998; Shaw et al., 2001; Yokochi and Marty, 2004) have used $S_{He}/S_{Ne}$ = 2 based on experiments in basaltic magmas at 1573–1623 K (Jambon et al., 1986; Lux, 1987). To investigate whether $S_{He}/S_{Ne}$ of 2 is applicable to discussions of a magma ocean, we used the ionic porosity model of noble gas solubility (Carroll and Stolper, 1993; Iacono-Marziano et al., 2010) to calculate $S_{He}/S_{Ne}$ for a bulk silicate Earth (pyrolite) magma composition (McDonough and Sun, 1995). We obtain $S_{He}/S_{Ne}$ of 1.96 in a magma ocean at 2053 K, the peridotite liquidus temperature at 1 bar (Katz et al., 2003), noting that the surface of a magma ocean should be at least this hot initially. Because $S_{He}/S_{Ne}$ decreases at higher temperatures (Fig. 5) and other proposed compositions of the bulk silicate Earth (Javoy et al., 2010; Palme and O'Neill, 2003) also yield $S_{He}/S_{Ne}$ <2, we adopt a value of 2, noting that the relevant value may be lower and not constant.

In the following discussion, we outline a possible chronology that includes at least three major accretionary events terminating with the Moon-forming giant impact: (i) ingassing of nebular gases into a magma ocean on the proto-Earth (Fig. 6a); (ii) loss of the nebular atmosphere, production of a partial mantle magma ocean by a giant impact, and subsequent outgassing to produce a secondary atmosphere (Fig. 6b); (iii) loss of the secondary atmosphere and generation and outgassing of another partial mantle magma ocean likely associated with the Moon-forming giant impact (Fig. 6c). This highly simplified chronology may be thought of as a minimum sequence of events that explains the main observation of a preserved low $^{3}He/^{22}Ne_m$ ratio in the deep plume mantle and a high $^{3}He/^{22}Ne_m$ ratio in the MORB mantle.

*6.1 Magma ocean ingassing and the OIB mantle*

Ingassing of nebular gas from a gravitationally accreted nebular atmosphere into a magma ocean has been proposed as the mechanism responsible for acquisition of terrestrial He and Ne (Harper and Jacobsen, 1996; Mizuno et al., 1980; Porcelli et al., 2001; Yokochi and Marty, 2004). The observation of solar-like $^{20}Ne/^{22}Ne$ ratios (>12.9) in primitive OIBs supports the presence of nebular He and Ne in the deep mantle (Mukhopadhyay, 2012; Yokochi and Marty, 2004). During nebular ingassing into a magma ocean, the $^{3}He/^{22}Ne$ ratio of the magma ocean would be fractionated from the nebular value of ~1.5 by $S_{He}/S_{Ne}$ of ≤2 to values of ~2.3–3 (Fig. 6a). Thus, the solar-like $^{20}Ne/^{22}Ne$ ratios and the $^{3}He/^{22}Ne$ ratios observed in plumes with the most primitive $^{21}Ne/^{22}Ne$ ratios (Fig. 3) are strong evidence for nebular ingassing.

The ingassing hypothesis is consistent with the timescales for planet formation and dissipation of nebular gases. The median timescale of dissipation for nebular gas is ~3 Ma, but nebular gas may be present for up to 10–12 Ma (Wyatt, 2008). Mars reached approximately half its present size in 1.8 ± 1 Ma (Dauphas and Pourmand, 2011) and the mean age of Earth's accretion is around 11 ± 1 Ma (Yin et al., 2002). Embryos of at least the mass of Mars are required for gravitational capture of substantial amounts of nebular gases (Hayashi et al., 1979). We suggest that the proto-Earth, poorly constrained to between 10% to ~60% of the present mass of Earth, gravitationally captured a nebular atmosphere that equilibrated with a magma ocean. This early magma ocean could have been produced through early radiogenic heating, the energy of accretion, and the blanketing effect of the massive nebular atmosphere, and may have occurred on multiple embryos that later accreted to form the Earth.

*6.2 Multiple magma oceans and the depleted mantle*

After the initial episode of ingassing, the $^3$He/$^{22}$Ne ratio of the shallower portion of the mantle increased to ≥10 while the deep mantle preserved values of ~2.3–3. If the origin of noble gases in the shallower mantle is meteoritic as opposed to solar (section 4), the required fractionation is even higher (Fig. 3). Because $S_{He}/S_{Ne}$ is at most 2 (Fig. 5), multiple episodes of magma ocean outgassing are required to increase the mantle $^3$He/$^{22}$Ne ratio from the value preserved in primitive plumes of at most 3 to the value in the depleted mantle of at least 10.

Magma ocean outgassing of the whole Earth was proposed by Honda and McDougall (1998) as a mechanism to increase the mantle $^3$He/$^{22}$Ne ratio. They derived an average mantle $^3$He/$^{22}$Ne ratio of 7.7 ± 2.6 from various MORBs and OIBs, and noted that this value is twice the solar value of 3.8. The authors argued that because $S_{He}/S_{Ne}$ in basaltic liquids is 2, the mantle's apparent 2× fractionation from the solar nebular value could be explained by solubility-controlled outgassing of a magma ocean. They also noted that the MORB average seemed higher than the OIB average, but based their arguments on the combined average. We argue that the average $^3$He/$^{22}$Ne ratio of the mantle is not particularly meaningful in the discussion of the extent of magma ocean degassing as the depleted mantle and the primordial reservoir clearly have different $^3$He/$^{22}$Ne$_m$ ratios (Fig. 3). Furthermore, the solar value of 3.8 is an unlikely primordial value because the solar nebula $^3$He/$^{22}$Ne ratio is 1.46 ± 0.06 and implanted solar gas has a $^3$He/$^{22}$Ne ratio of 0.9 ± 0.1 (Fig. 3).

We propose that after loss of the nebular atmosphere and during the giant impact phase of terrestrial growth, at least two separate partial mantle magma ocean episodes raised the $^3$He/$^{22}$Ne ratio of the shallower mantle to ≥10 (a modification of the hypothesis proposed by Honda and McDougall, 1998). The last of the magma ocean episodes would have been associated with the Moon-forming giant impact. Equilibrium outgassing of a magma ocean can increase the magmatic $^3$He/$^{22}$Ne ratio by up to a factor of $S_{He}/S_{Ne}$, which is ≤2 (Fig. 5). Loss of the nebular atmosphere, either before or during a giant impact, followed by solubility-controlled outgassing from the impact-generated magma ocean, raised the $^3$He/$^{22}$Ne ratio from at most 2.3–3 to as high as 5–6 (Fig. 6b). A minimum of one additional episode of atmospheric loss and impact-induced magma ocean outgassing is then required to achieve the $^3$He/$^{22}$Ne ratio of ≥10 that characterizes the depleted mantle (Fig. 6c).

Partial or complete loss of the preexisting atmosphere prior to magma ocean generation is a requirement to drive the $^3$He/$^{22}$Ne fractionation in the magma ocean. If an atmosphere that had equilibrated with a previous magma ocean remained intact during a giant impact, the new magma ocean would not outgas and fractionate $^3$He from $^{22}$Ne because it would already be in equilibrium with the atmosphere. Complete loss would lead to the highest degree of He/Ne fractionation, up to $S_{He}/S_{Ne}$, whereas partial loss would lead to fractionation less than $S_{He}/S_{Ne}$ ratio. Therefore, if atmospheric loss between magma ocean episodes were incomplete, more than the minimum of two magma ocean outgassing events would be required to attain $^3He/^{22}Ne$ ratios of ≥10.

Hydrodynamic escape and impact erosion could provide the mechanisms for the atmospheric loss required by the $^3He/^{22}Ne$ observations. The $H_2$-rich primary nebular atmosphere captured by embryos or the proto-Earth may have been partially or substantially lost through hydrodynamic escape after the nebula dissipated (e.g., Pepin, 1991, see section 7.1). However, secondary (outgassed) atmospheres are unlikely to be $H_2$-rich even if core formation is ongoing (Hirschmann, 2012) and consequently, hydrodynamic escape may not be an important process for loss of outgassed atmospheres. Furthermore, the process would result in chemical fractionations that are not observed (Marty, 2012; Sharp and Draper, 2013). We therefore propose that loss of secondary atmospheres was driven by giant impacts during terrestrial accretion. Although numerical simulations suggest that atmospheres are not necessarily lost during a giant impact (Genda and Abe, 2003), atmospheric loss becomes more efficient if the planet has an ocean (Genda and Abe, 2005) and/or if the planet is rotating quickly (Lock and Stewart, 2013; Stewart and Mukhopadhyay, 2013). Irrespective of the precise mechanism for atmospheric loss, we reiterate that atmospheric loss is a requirement for driving $^3He/^{22}Ne$ fractionation in a magma ocean (Fig. 6).

These atmospheric loss and magma ocean episodes would have occurred as planetary embryos started colliding with each other following dissipation of the nebula. While the last of these magma oceans must have occurred on the Earth in association with the Moon-forming impact, the previous magma ocean outgassing episode(s) may have happened on the planetary embryos that collided with the proto-Earth. However, giant impacts were prevalent during accretion (e.g., O'Brien et al., 2006; Walsh et al., 2011) and two giant impacts on the Earth itself are implied in a recent hypothesis for the formation of the Moon (Ćuk and Stewart, 2012). To obtain a lunar disk with the same bulk composition as the Earth, the hypothesis posits a giant impact onto a fast-spinning Earth. To spin the Earth up requires an additional giant impact prior to the Moon-forming impact.

Our proposed scenario explains the increase in the mantle $^3$He/$^{22}$Ne ratio through equilibrium outgassing of at least two partial mantle magma oceans. If there were disequilibrium degassing, the fractionation between He and Ne would be suppressed (e.g., Gonnermann and Mukhopadhyay, 2007; Paonita and Martelli, 2007). Consequently, the $^3$He/$^{22}$Ne ratio in the magma ocean would be fractionated by less than $S_{He}/S_{Ne}$, requiring additional magma ocean outgassing and atmospheric loss events. In volcanic systems, open-system degassing can fractionate elemental ratios beyond the solubility ratio (e.g., Moreira and Sarda, 2000) and open system magma ocean degassing has been proposed (Shaw et al., 2001). Open-system degassing of a magma ocean implies that the degassed gases are immediately lost from the atmosphere, even though Ne is gravitationally bound. Such gas loss could be driven by hydrodynamic escape during outgassing of the magma ocean, perhaps precluding the need for multiple outgassing episodes. To test this possibility, we performed a coupled magma ocean outgassing/hydrodynamic escape calculation. In our calculation, EUV flux from the young Sun drives $H_2$ loss that in turn lifts He and Ne from the atmosphere. Since the atmosphere and magma ocean remain in equilibrium, atmospheric loss leads to further outgassing from the magma ocean (Fig 7; equations in supplementary material). We find that despite the open-system nature of outgassing, hydrodynamic escape is significantly more efficient at removing He from the atmosphere than Ne, such that the $^3$He/$^{22}$Ne ratios of the atmosphere and magma ocean decrease (Fig. 7). Hence, we conclude that two outgassing episodes are the *minimum* required to explain the fractionated $^3$He/$^{22}$Ne of the depleted mantle.

# 7. Implications for Earth's accretional history

The observation of large differences in $^3$He/$^{22}$Ne ratios in the mantle and the requirement for atmospheric loss in our magma ocean outgassing hypothesis have significant implications for the terrestrial inventories of volatile elements, early mantle mixing, and the origin of mantle heterogeneities.

*7.1 The terrestrial volatile budget*

Magma ocean outgassing can efficiently transfer volatiles (e.g., $N_2$, $H_2O$, and $CO_2$) from the mantle to the exosphere (Elkins-Tanton, 2008; Zahnle et al., 2007). Consequently, the mantle was dried by giant impacts and the terrestrial volatile budget was concentrated at the surface *prior* to the Moon-forming giant impact. As a result, the volatiles would have been susceptible to loss through atmospheric blow-off during the Moon-forming giant impact.

Multiple atmospheric loss and magma ocean outgassing episodes are compatible with the He and Ne abundances inferred for the present-day mantle, although these abundances are poorly known (see supplementary material). However, the atmospheric loss episodes, which are necessary to explain $^3$He/$^{22}$Ne fractionation during magma ocean outgassing, may also provide explanations for other curious features of the terrestrial volatile budget.

The combined silicate mantle and exosphere of the Earth has a low C/H ratio and extremely low N/H ratio compared to CI chondrites (Fig. 8; Halliday, 2013; Hirschmann and Dasgupta, 2009; Marty, 1995), which is surprising considering the Earth's chondritic-like C, H, and N isotopic ratios (e.g., Alexander et al., 2012; Marty, 2012). The depletion of carbon has been linked to atmospheric loss or to partitioning of C into the core (Hirschmann and Dasgupta, 2009). Likewise, Marty (2012) argued that nitrogen depletion is due to sequestration of N in the core. However, the depletion of N by core formation is unlikely because it would be accompanied by a significantly greater depletion of C; $D_N^{metal/silicate} \leq 10$ (Katik et al., 2011), whereas $D_C^{metal/silicate}$ may be >1000 (Dasgupta et al. 2013). We suggest that the observed moderate depletion in C and extreme depletion in N can be explained if the Earth had a water ocean prior to the last giant impact. The presence of liquid water on an accreting Earth is likely if the time between giant impacts is >$10^7$ years (Elkins-Tanton, 2008; Zahnle et al., 2007). A giant impact would fractionate the C/H and N/H ratios as it would preferentially remove the atmosphere compared to the ocean (Genda and Abe, 2005). Because C can form bicarbonate and carbonate ions, a significantly larger proportion of C would be in the ocean (or in the crust) compared to N. Therefore, preferential loss of the atmosphere would lead to a large decrease in the N/H ratio compared to a modest decrease in the C/H ratio (Fig. 8).

An atmospheric loss event that also involved partial loss of an ocean could explain why the heavy halogens (Cl, Br) are depleted in the Earth relative to fluorine and other highly volatile elements (Fig 8; McDonough, 2003; Sharp and Draper, 2013). This is another peculiar aspect of terrestrial volatile abundances (Marty, 2012), especially considering that the Earth has a chondritic Cl isotope ratio (Sharp and Draper, 2013). F is highly soluble in magmas compared to Cl and Br (e.g. molar $S_F/S_{Cl}$ = ~4 where $S$ is the melt/vapor solubility coefficient; Auippa, 2009). Consequently, during magma ocean outgassing Cl and Br would be outgassed while F would be preferentially retained in the magma ocean. Removal of the atmosphere or partial removal of an ocean during a giant impact could, therefore, explain the depletion of the heavy halogens on the Earth with respect to F.

We argue that the relative proportions of major volatiles (H, C, N, halogens) on Earth record the violent events during accretion. Thus, most of Earth's water, and possibly the other major volatiles, were acquired prior to and during the giant impact phase of terrestrial accretion (also see Halliday, 2013; Saal et al., 2013). The observed fractionations in the major volatiles (Fig. 8) could have been larger in the immediate aftermath of the Moon-forming giant impact, as the late veneer also contributed volatiles to Earth (e.g., Albarède, 2009; Wänke, 1981). However, the late veneer contribution was not sufficiently large to overprint the volatile characteristics acquired during the main stages of Earth's accretion (Halliday, 2013).

*7.2 Preservation of early-formed heterogeneity and depth of the magma oceans*

The observation of different $^3$He/$^{22}$Ne ratios in the OIB and MORB mantle reservoirs requires the formation and preservation of two distinct mantle domains during accretion. This implies that the later magma oceans, including the one created by the Moon-forming giant impact, did not homogenize the whole mantle. Because the timescale for turnover of a turbulently convecting magma ocean may be as short as a few weeks (Elkins-Tanton, 2008; Pahlevan and Stevenson, 2007), a magma ocean would be expected to be chemically homogeneous. The deep mantle could preserve a low $^3$He/$^{22}$Ne ratio if the later giant impacts did not melt the whole mantle (Fig. 6b,c). Regions in the deep mantle could escape melting because the distribution of energy during a giant impact is heterogeneous (Solomatov, 2000). Along with $^3$He/$^{22}$Ne ratios, the differences in $^{20}$Ne/$^{22}$Ne (Mukhopadhyay, 2012; Yokochi and Marty, 2004) and $^{129}$Xe/$^{130}$Xe ratios between the deeper plume mantle and shallower depleted MORB mantle (Mukhopadhyay, 2012; Parai et al., 2012; Pető et al., 2013; Tucker et al., 2012) support the notion that the mantle was never completely homogenized during or since Earth's accretion.

*7.3 Formation of the LLSVPs*

Large Low Shear-Velocity Provinces (LLSVPs), two large antipodal seismically imaged structures at the base of the mantle, have been linked to OIB volcanism (e.g., Burke et al., 2008; Thorne et al., 2004; Wen and Anderson, 1997). LLSVPs are often associated with crystal fractionation from a basal magma ocean (e.g., Coltice et al., 2011; Labrosse et al., 2007) and Coltice et al. (2011) have suggested that the primitive He and Ne signature in OIBs reflects the signature of a basal magma ocean. However, the primitive reservoir with a low $^3$He/$^{22}$Ne ratio cannot have its origin in a basal magma ocean produced by the last giant impact (the Moon-forming giant impact) because that magma ocean must have produced the high $^3$He/$^{22}$Ne ratio now sampled in the depleted mantle (Section 6.2). A basal magma ocean sequestering low $^3$He/$^{22}$Ne ratios in the deep mantle could have formed from a previous whole-mantle magma ocean. In this case, subsequent giant impacts including the Moon-forming giant impact must not have disrupted the basal magma ocean, whose lifetime is >1 Gyr (Labrosse et al., 2007), or else this reservoir of primitive material would not remain intact. In summary, LLSVPs may have formed as crystallization products of previous magma oceans and survived the Moon-forming giant impact. Otherwise, their formation is unrelated to magma oceans.

# 8. Conclusions

We observe $^3$He/$^{22}$Ne$_m$ ratios in equatorial Atlantic MORBs to be correlated with Ne, Pb,

and Nd isotopic ratios, which allows us to constrain the depleted mantle $^3$He/$^{22}$Ne ratio to be $\geq 10$. In contrast, the deep primitive reservoir sampled by plumes has a ratio of ~2.3 to 3. We find that processes associated with the long-term plate tectonic cycle, such as partial melting and recycling of plates, are incapable of changing the mantle $^3$He/$^{22}$Ne ratio appreciably in 4.5 Ga.

Solubility-controlled fractionation associated with ingassing of a gravitationally captured nebular atmosphere on the proto-Earth raised the mantle $^3$He/$^{22}$Ne ratio from a primordial value of $\leq 1.5$ to that sampled in primitive plumes. Atmospheric loss and magma ocean outgassing associated with at least two separate giant impacts subsequently increased the mantle $^3$He/$^{22}$Ne ratio to $\geq 10$. These giant impacts, including the Moon-forming impact, likely did not melt the whole mantle or else the low $^3$He/$^{22}$Ne ratios sampled in primitive plumes could not be preserved. Atmospheric loss and magma ocean outgassing also provide an explanation for the lower than chondritic C/H and N/H ratios as well as Cl/F and Br/F ratios in the Earth.

**Acknowledgements:** We thank Marc Hirschmann, Simon Lock, Rita Parai, and Sarah Stewart for helpful discussions, and editor Tim Elliott and two anonymous reviewers for helpful comments. This work was supported by NSF grant OCE 0929193.

**Figures**

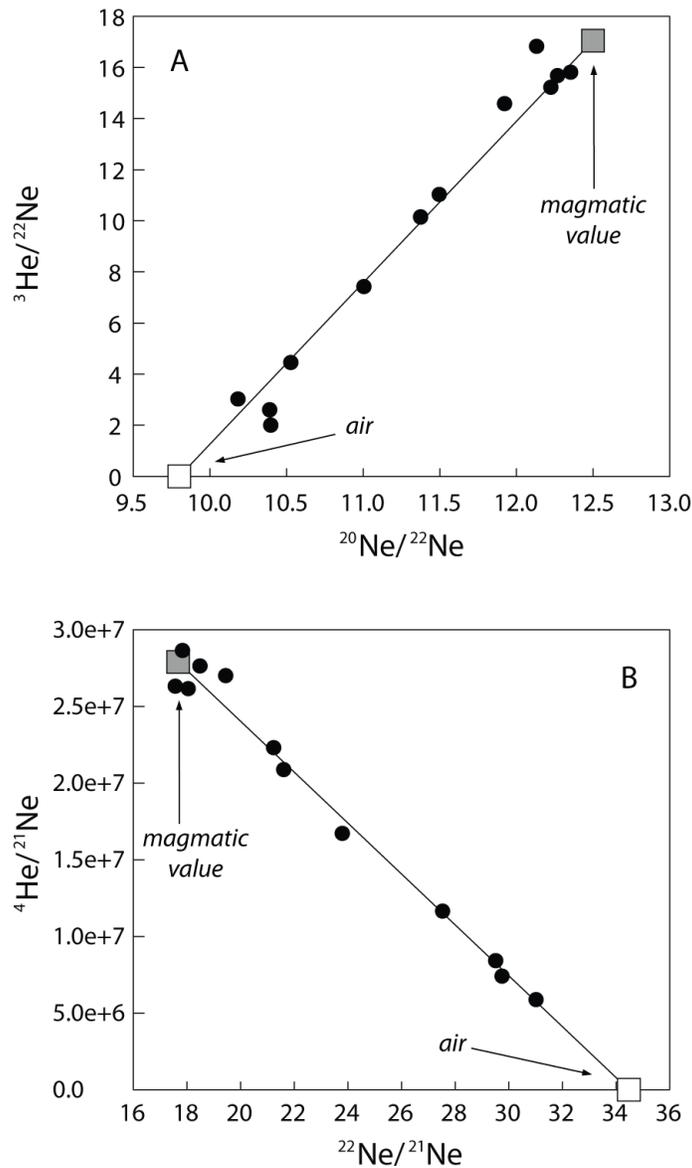

**Figure 1**: Example of correlation diagrams to determine the MORB source $^3$He/$^{22}$Ne ($^3$He/$^{22}$Ne$_m$) ratios. Each point represents a single crushing step of the equatorial Atlantic MORB sample RC2806 42D-7 (data presented in Tucker et al., 2012). Step-crushing releases gas trapped in vesicles that reflects magmatic gas variably contaminated with a post-eruptive atmospheric contaminant. Extrapolation of the (a) $^3$He/$^{22}$Ne-$^{20}$Ne/$^{22}$Ne and (b) $^4$He/$^{21}$Ne-$^{22}$Ne/$^{21}$Ne correlation trends to the uncontaminated mantle values of $^{20}$Ne/$^{22}$Ne = 12.5 and $^{22}$Ne/$^{21}$Ne (determined by correlation of $^{20}$Ne/$^{22}$Ne with $^{21}$Ne/$^{22}$Ne; Tucker et al., 2012) corrects for post-eruptive air contamination and establishes the magmatic $^3$He/$^{22}$Ne and $^4$He/$^{21}$Ne ratios. This magmatic $^3$He/$^{22}$Ne ratio is corrected for potential magmatic degassing though the degree to which the magmatic ($^4$He/$^{21}$Ne)* ratio is fractionated from the mantle ($^4$He/$^{21}$Ne)* production ratio (see supplementary material).

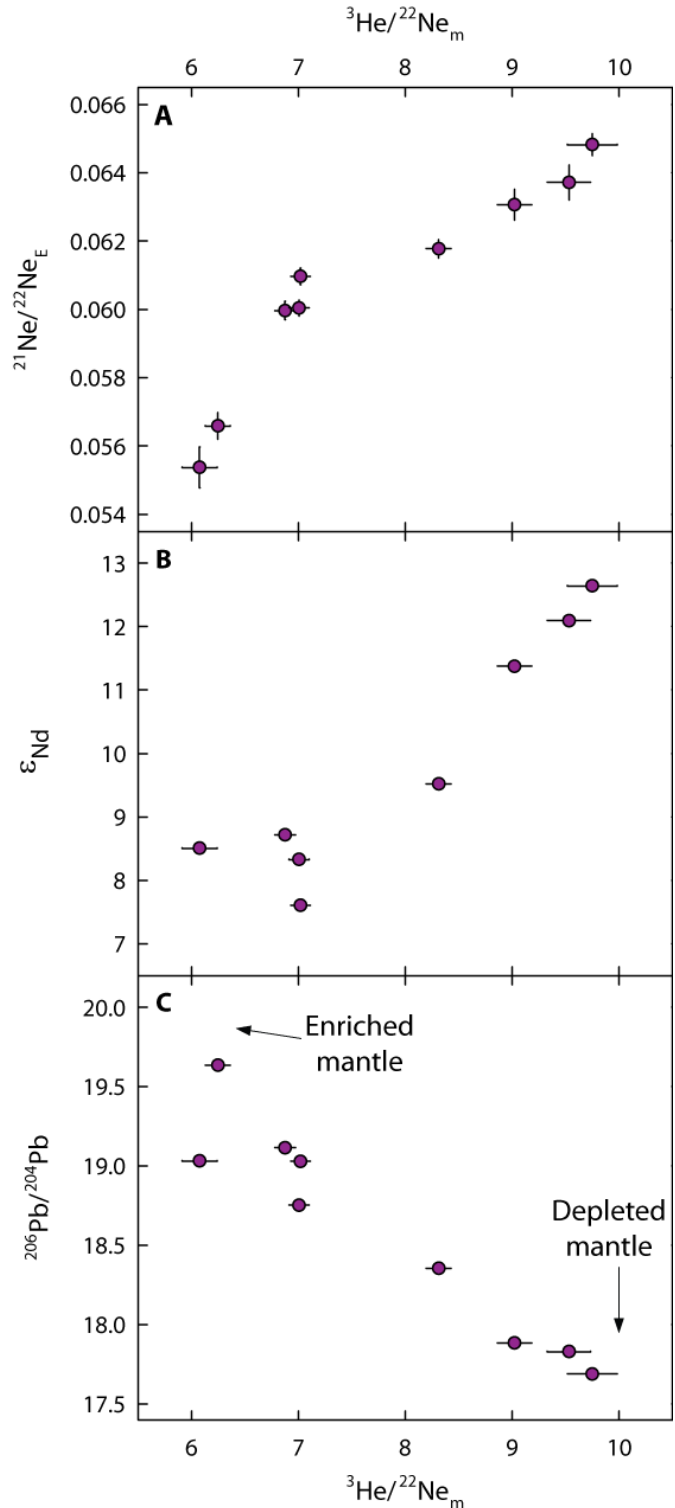

**Figure 2**: Correlation between $^3$He/$^{22}$Ne$_m$ and (a) $^{21}$Ne/$^{22}$Ne, (b) $\varepsilon_{Nd}$ where $\varepsilon$ is the part in $10^4$ deviation from the chondritic $^{143}$Nd/$^{144}$Nd value, and (c) $^{206}$Pb/$^{204}$Pb. These correlations establish the depleted mantle value to be at least 9.8 and indicate that the variability in MORB $^3$He/$^{22}$Ne$_m$ ratios is caused by recent mixing between a depleted mantle source and a more enriched plume source with a lower $^3$He/$^{22}$Ne ratio.

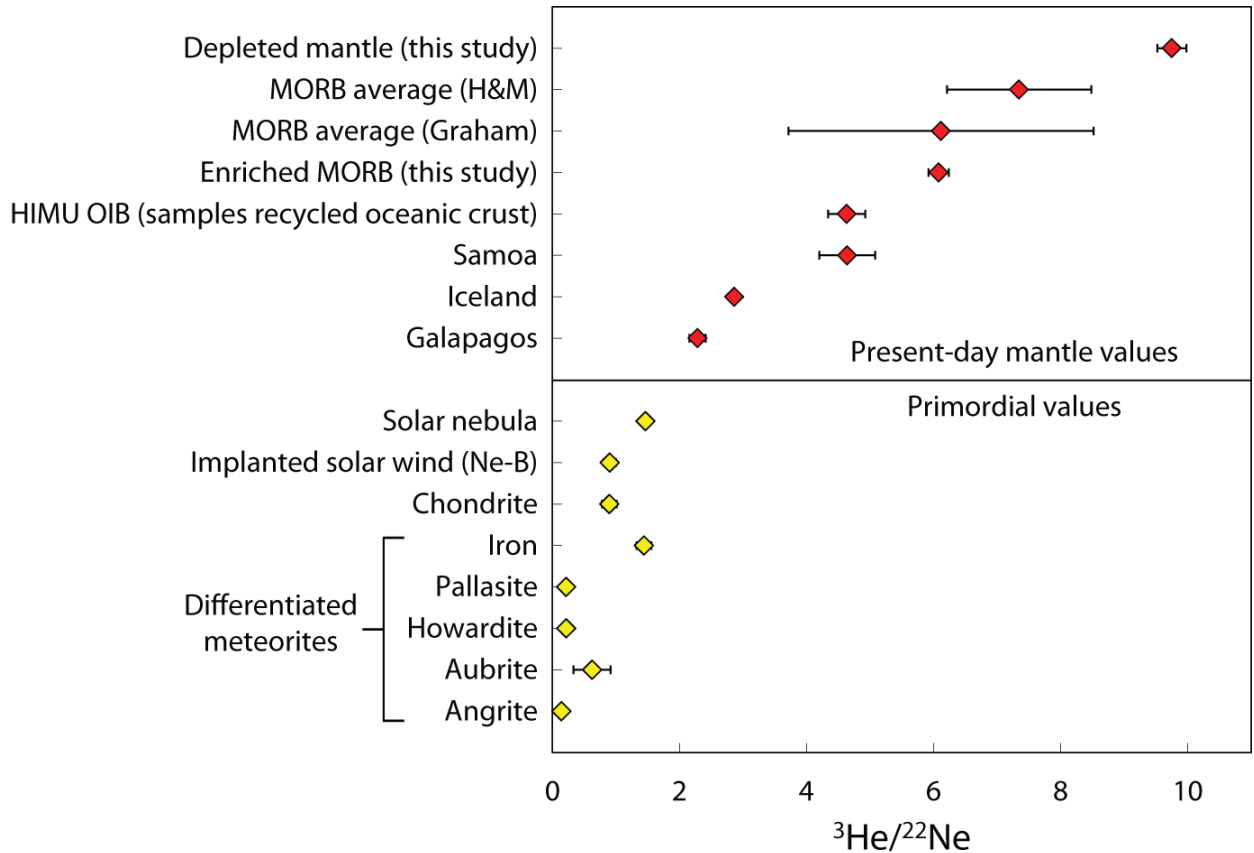

**Figure 3**: $^3$He/$^{22}$Ne$_m$ ratios in modern terrestrial reservoirs along with the $^3$He/$^{22}$Ne ratios in possible sources that may have contributed primordial He and Ne to the Earth. Note that the depleted mantle value is at least a factor of 6.5 higher than the value in possible primordial materials. MORBs also have higher values than the primitive reservoir sampled by OIBs at Galapagos and Iceland. $^3$He/$^{22}$Ne ratios for terrestrial reservoirs are computed by Method 1. Details of the calculations are given in the supplementary material. Data sources are as follows: MORB average (H&M) from Honda and McDougall (1998); MORB average (Graham) from Graham (2002); Mangaia from Parai et al. (2009); Samoa from Jackson et al. (2009); Iceland from Mukhopadhyay (2012); Galapagos from Kurz et al. (2009) and Raquin and Moreira (2009); solar nebula from Grimberg et al. (2006), Mahaffy et al. (1998), and Pepin et al. (2012); implanted solar wind from Raquin and Moreira (2009); chondrite from Ott (2002); iron (Washington County; unclassified) from Becker and Pepin (1984); pallasite (Brenham; olivine) from Mathew and Begemann (1997); howardite (Kapoeta and Jodzie) from Mazor and Anders (1967); aubrites (various) from Lorenzetti et al. (2003); angrite (D'Obrigny) from Busemann et al. (2006). The chondrite, pallasite, howardite, and angrite values are maximum values (see supplementary material).

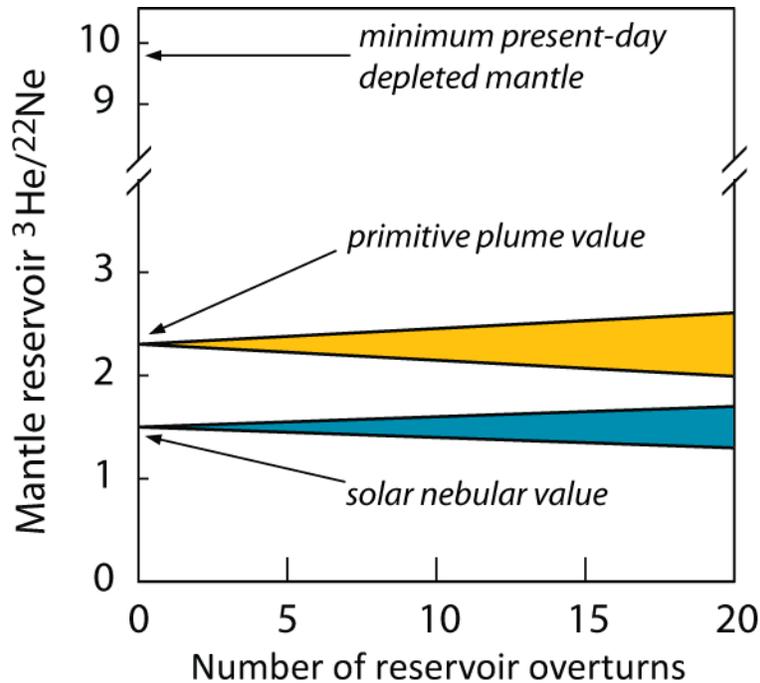

**Figure 4**: Model calculation showing the $^3$He/$^{22}$Ne evolution of a hypothetical mantle reservoir by plate tectonic cycling. The $^3$He/$^{22}$Ne ratio evolves as a function of the number of reservoir masses processed through partial melting. The Earth's mantle has likely experienced <9 overturns in 4.5 Ga (Coltice et al., 2009; Gonnermann and Mukhopadhyay, 2009), so plate tectonic cycling cannot significantly change the mantle $^3$He/$^{22}$Ne ratio. This result is not sensitive to the initial value chosen; starting from the solar nebula value of ~1.5 the mantle cannot evolve to the primitive reservoir values of ~2.3–3 sampled by OIBs, and starting from 2.3, the mantle cannot evolve to the value of ≥10 observed in the depleted mantle. The plot shows the range of possible $^3$He/$^{22}$Ne evolution by combining the two extremes (1σ) in the He and Ne partition coefficients of Brooker et al. (2003) and Heber et al. (2007) (Table 2). For example, we pair the +1σ in the He partition coefficient with the −1σ in the Ne partition coefficient. We note that if the average degree of melting is greater than 5%, or the melt is extracted more efficiently than in batch melting, or the mantle has experienced fewer than 20 overturns, the overall change in the mantle $^3$He/$^{22}$Ne ratio would be less than that shown in the figure. Additionally, if recycled crust carries Ne but not He, then the mantle $^3$He/$^{22}$Ne would decrease over time.

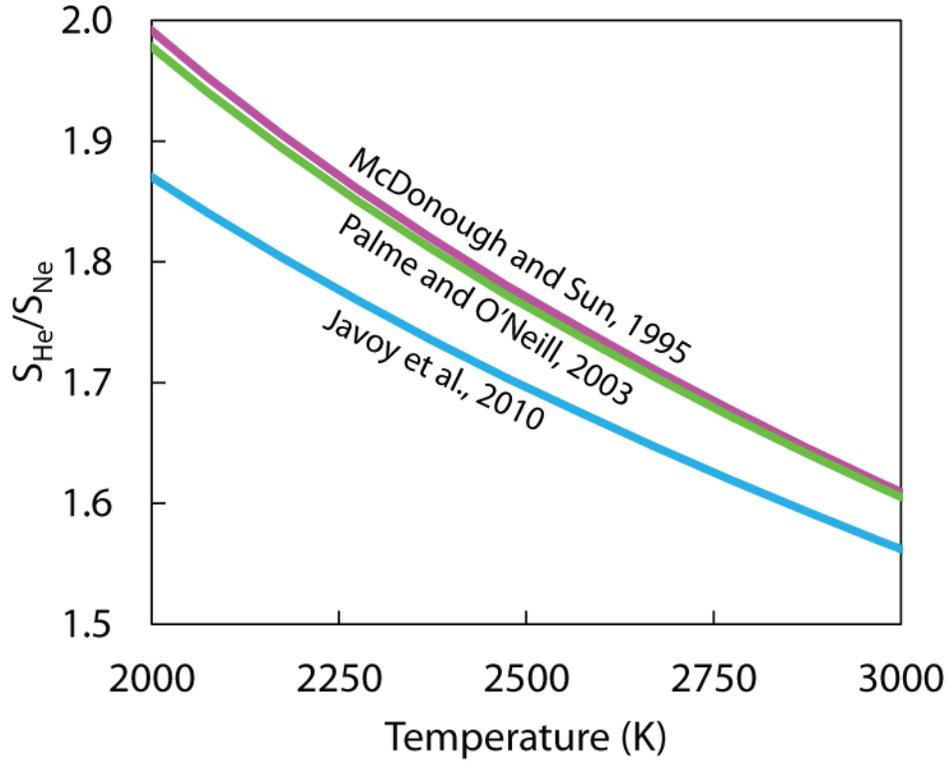

**Figure 5**: The solubility ratio of He to Ne ($S_{He}/S_{Ne}$) corresponding to three magma ocean compositions, based on three proposed bulk silicate Earth compositions, as a function of temperature. Solubility $S$ (Henry's constant) is the melt/vapor partition coefficient, computed from the solubility model of Iacono-Marziano et al. (2010). $S_{He}/S_{Ne}$ is at most 1.96, corresponding to the peridotite liquidus temperature of 2053 K at 1 bar.

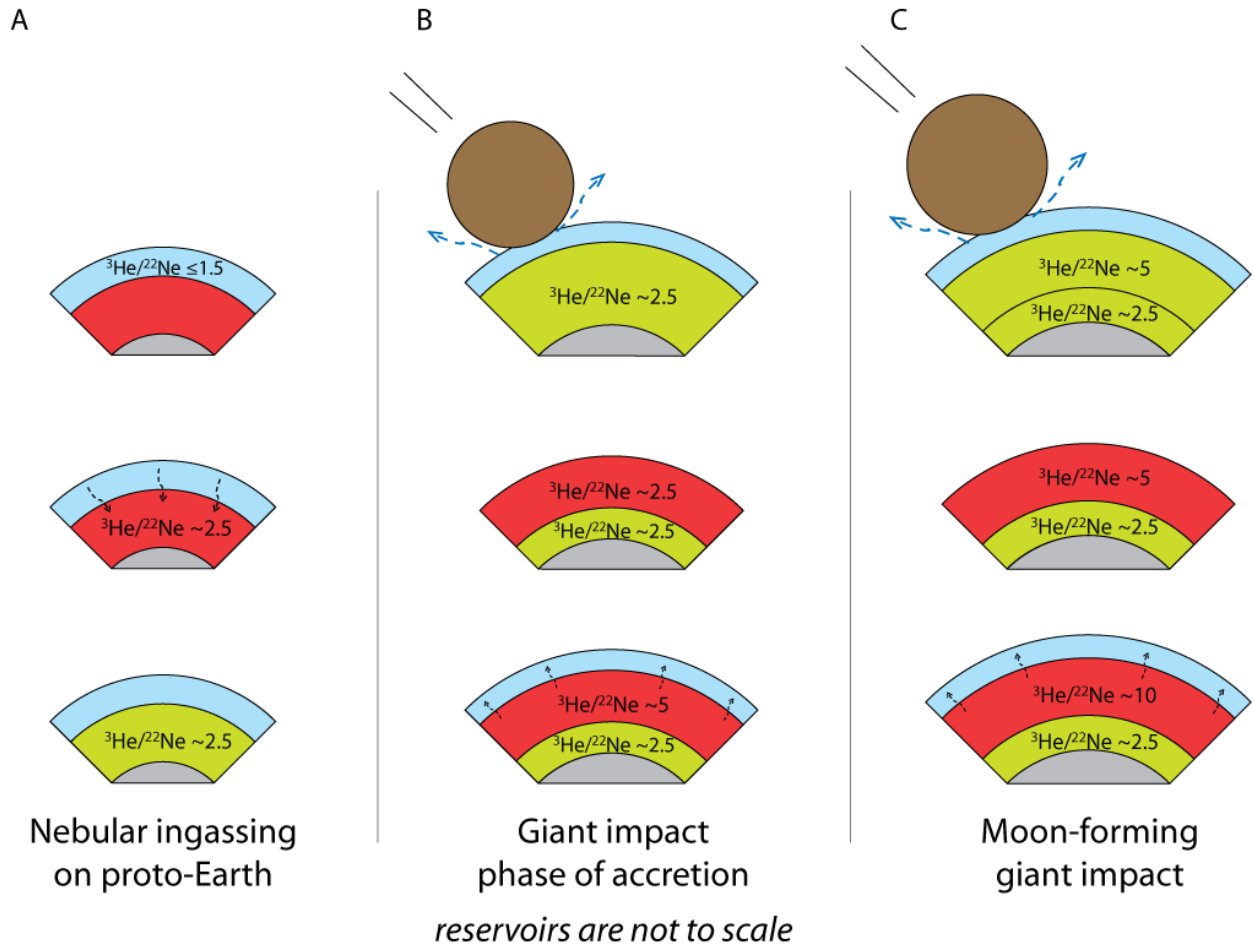

**Figure 6**: A proposed conservative chronology of events to explain the highly fractionated $^3$He/$^{22}$Ne ratio of the depleted mantle. (a) Nebular ingassing. A gravitationally accreted nebular atmosphere surrounds the molten proto-Earth. Gases are dissolved into the magma ocean and fractionated by up to their solubility ratio. The magma ocean solidifies, and the mantle inherits the fractioned $^3$He/$^{22}$Ne ratio. Nebular ingassing probably happened on multiple embryos that later collided to form the Earth. (b) Magma ocean outgassing. A giant impact (or hydrodynamic escape) leads to loss of the existing atmosphere and produces a new partial mantle magma ocean, preserving the low $^3$He/$^{22}$Ne reservoir sampled by deep mantle plumes. This magma ocean outgasses, forming a secondary atmosphere, and leaving residual He and Ne fractionated. (c) Moon-forming giant impact. A second giant impact blows off the existing secondary atmosphere and induces a new partial mantle magma ocean. This magma ocean outgasses, forming a new secondary atmosphere, and leaving He and Ne fractionated. Because the amount of $^3$He/$^{22}$Ne fractionation in each step is at most a factor of 2, our data require at least two giant impact-induced magma ocean outgassing episodes to achieve a $^3$He/$^{22}$Ne ratio ≥10.

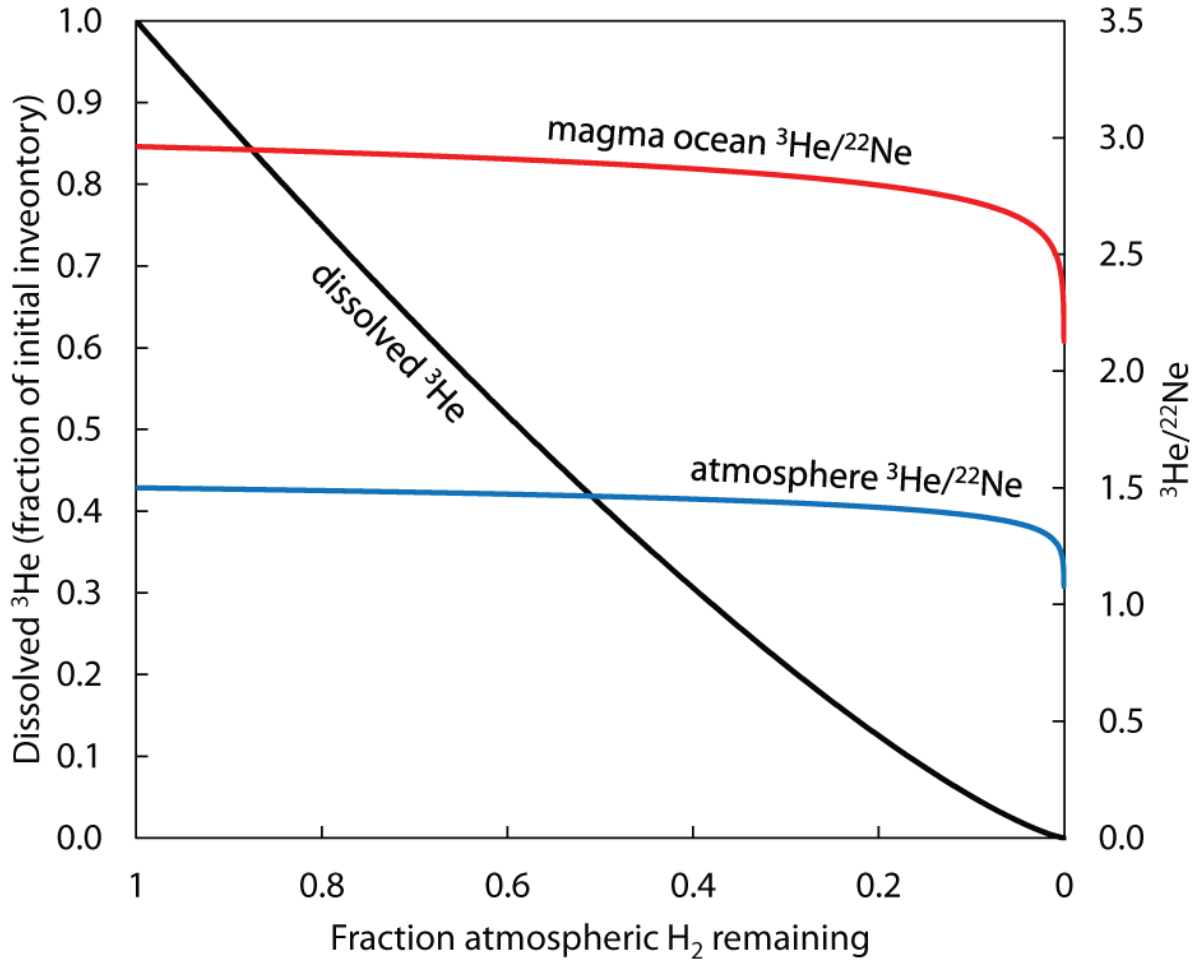

**Figure 7:** Coupled magma ocean outgassing and hydrodynamic escape model. Plotted are the dissolved magmatic $^3$He abundance (left axis) and magma ocean and atmospheric $^3$He/$^{22}$Ne ratios (right axis) as a function of atmospheric H$_2$ inventory, a proxy for time. In this model, solar EUV radiation, 100× the present value (Ribas et al., 2005), drives H$_2$ loss from a nebular atmosphere, which lifts He and Ne from the atmosphere. The underlying magma ocean is always assumed to be in equilibrium with the atmospheric composition above it so the maximum amount of outgassing can occur. In the calculation shown we assume a proto-Earth of 0.3 × Earth mass, an initial surface pressure of 30 bar, surface temperature of 2000 K, and H$_2$ escape flux that decays with a time constant of 90 Myr (Pepin, 1991); the corresponding initial crossover mass are 541 and 762 amu for He and Ne, respectively. Equations are given in the supplementary material. We note that the $^3$He/$^{22}$Ne ratios in the atmosphere and magma ocean decrease regardless of the particular parameters used, and therefore conclude that open-system degassing accommodated by hydrodynamic escape cannot raise the mantle $^3$He/$^{22}$Ne ratio.

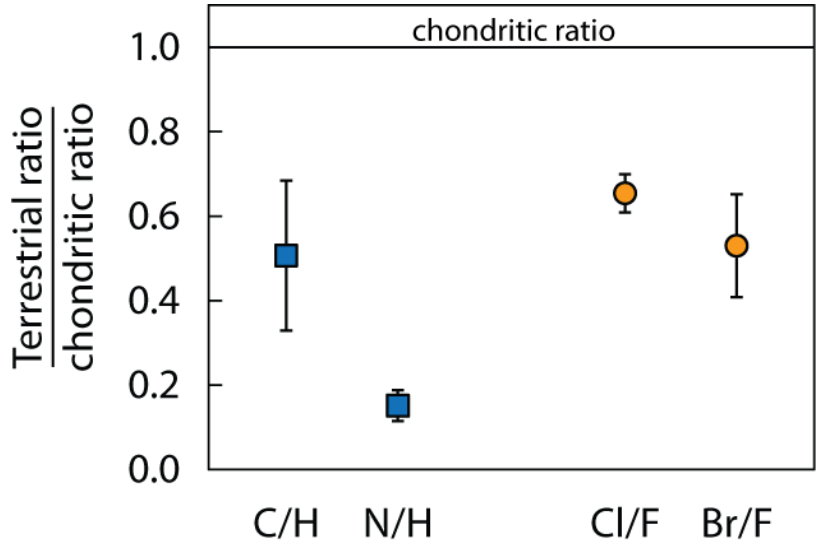

**Figure 8**: Relative terrestrial abundances of C, N, H, and halogens. The relative depletions of C and N with respect to H can be explained by atmospheric loss during a giant impact if H is condensed as a water ocean as atmosphere would be lost in preference to liquid water. The greater depletion of N could reflect some C retention as bicarbonate and carbonate ions in the ocean and/or carbonated crust. The figure also illustrates the relative depletion of Cl and Br with respect to F, a surprising feature of the terrestrial volatile pattern. At least two episodes of magma ocean outgassing, as inferred from the $^3$He/$^{22}$Ne data, would have outgassed Cl and Br and concentrated them in the exosphere prior to the last giant impact. The higher solubility of F in magmas would cause F to be preferentially retained in the mantle. Loss of the atmosphere or partial loss of the hydrosphere during a giant impact may explain the relative depletion Cl and Br compared to F. Terrestrial and chondritic C, N, and H abundances from Halliday (2013); terrestrial halogen abundances from McDonough (2003); chondritic halogen abundances from Lodders (2003).

**Table 1: MORB source $^3$He/$^{22}$Ne ratios**

| | $^4$He/$^3$He | $^{21}$Ne/$^{22}$Ne$_E$ | $^3$He/$^{22}$Ne$_m$ Method 1 | | $^3$He/$^{22}$Ne$_m$ Method 2 | |
|---|---|---|---|---|---|---|
| ($^4$He/$^{21}$Ne)* production ratio | | | $2.2 \times 10^{7\dagger}$ | $2.8 \times 10^{7\ddagger}$ | $2.2 \times 10^{7\dagger}$ | $2.8 \times 10^{7\ddagger}$ |
| *Depleted MORBs* | | | | | | |
| EN061 2D | 80800 ± 1080 | 0.0637 ± 0.0005 | 9.53 ± 0.20$^\S$ | 12.14 ± 0.26 | 9.09 ± 0.70 | 11.57 ± 0.90 |
| EN061 4D | 81700 ± 1670 | 0.0648 ± 0.0003 | 9.75 ± 0.23 | 12.41 ± 0.30 | 9.66 ± 0.19 | 12.30 ± 0.24 |
| RC2806 2D-1 | 83500 ± 830 | 0.0631 ± 0.0004 | 9.02 ± 0.16 | 11.48 ± 0.20 | 9.14 ± 0.22 | 11.63 ± 0.28 |
| RC2806 3D-2 | 86700 ± 870 | 0.0618 ± 0.0003 | 8.31 ± 0.12 | 10.58 ± 0.15 | 8.52 ± 0.25 | 10.85 ± 0.32 |
| *HIMU-type MORBs* | | | | | | |
| RC2806 24D-1 | 93200 ± 930 | 0.0554 ± 0.0006 | 6.08 ± 0.16 | 7.73 ± 0.21 | 6.21 ± 0.28 | 7.90 ± 0.35 |
| RC2806 42D-7 | 95100 ± 950 | 0.0566 ± 0.0004 | 6.25 ± 0.12 | 7.95 ± 0.15 | 6.42 ± 0.19 | 8.17 ± 0.24 |
| RC2806 59D-1 | 96300 ± 960 | 0.0600 ± 0.0002 | 7.01 ± 0.09 | 8.92 ± 0.12 | 6.91 ± 0.16 | 8.79 ± 0.20 |
| RC2806 48D-9 | 97700 ± 980 | 0.0600 ± 0.0003 | 6.88 ± 0.10 | 8.75 ± 0.12 | 6.89 ± 0.86 | 8.77 ± 1.09 |
| RC2806 57D-1 | 99000 ± 990 | 0.0610 ± 0.0002 | 7.02 ± 0.09 | 8.94 ± 0.12 | 7.00 ± 0.26 | 8.91 ± 0.33 |
| *MORB averages* | | | | | | |
| Honda and McDougall (1998)[1] | | | 7.35 ± 1.13 | | | |
| Graham (2002)[2] | | | 6.12 ± 2.40 | | | |

† Yatsevich and Honda (1997)

‡ Leya and Wieler (1999)

§ Uncertainties in $^3$He/$^{22}$Ne$_m$ (source $^3$He/$^{22}$Ne ratios) are propagated from uncertainties in $^4$He/$^3$He and $^{21}$Ne/$^{22}$Ne$_E$ reported in Tucker et al. (2012) (Method 1) and additionally from $^3$He/$^{22}$Ne$_E$ and $^4$He/$^{21}$Ne$_E$ (Method 2). The $^{21}$Ne/$^{22}$Ne$_E$ is the mantle $^{21}$Ne/$^{22}$Ne ratio obtained by correcting the measured isotopic ratio for syn- to post-eruptive air contamination (Tucker et al., 2012).

1. Recalculated from original data compiled in Honda and McDougall (1998). Details given in supplement.

2. Reported MORB average in Graham (2002) recalculated to $^{20}$Ne/$^{22}$Ne = 12.5. Details given in supplementary material.

**Table 2: He and Ne partition coefficients**

|  |  | He | 1σ | Ne | 1σ |
|---|---|---|---|---|---|
| Olivine | Heber 2007 | 0.00017 | 0.00013 | 0.00007 | 0.00007 |
|  | Jackson 2013[†] | 0.00017 |  |  |  |
| Clinopyroxene | Heber 2007 | 0.00020 | 0.00019 | 0.00041 | 0.00035 |
|  | Brooker 2003 |  |  | 0.00039 | 0.00019 |
|  | Jackson 2013 | 0.00013 |  |  |  |
| Orthopyroxene | Jackson 2013 | 0.00024 |  | [‡] |  |
| Spinel | Jackson 2013 | 0.00014* |  |  |  |

[†] Jackson et al. (2013) do not provide uncertainty estimates so their results are not included in our modeling. However, their reported values for olivine and clinopyroxene are within the 1σ range of previous studies.

[‡] Because the Ne partition coefficient in orthopyroxene is unconstrained, we assume it is similar to clinopyroxene based on the similar He partition coefficients for these two minerals.

* The reported value for spinel is within the 1σ range of other mantle minerals so explicit inclusion would not affect the results of our model.